# High power microwave system based on power combining and pulse compression of conventional klystrons

Zheng-Feng Xiong [1,2](熊正锋)   Huai-Bi Chen [1](陈怀壁)   Cheng Cheng [1](程诚)   Hui Ning [2](宁辉)   Chuan-Xiang Tang [1](唐传祥)

1 Department of Engineering Physics, Tsinghua University, Beijing 100084, China

2 Laboratory on Science and Technology of High Power Microwave, Northwest Institute of Nuclear Technology, Xi'an, Shaanxi 710024, China

**Abstract**: A high power microwave system based on power combining and pulse compression of conventional klystrons is introduced in this paper. This system mainly consists of pulse modulator, power combiner, driving source of klystrons and pulse compressor. A solid state induction modulator and pulse transformer were used to drive two 50 MW S-band klystrons with pulse widths 4 μs in parallel, after power combining and pulse compression, the tested peak power had reached about 210 MW with pulse widths nearly 400 ns at 25 Hz, while the experimental maximum output power was just limited by the power capacity of loads. This type of high power microwave system has widely application prospect in RF system of large scale particle accelerators, high power radar transmitters and high level electromagnetic environment generators.

## 1 Introduction

In the field of large scale particle accelerators, the RF peak power feed into the accelerator structure is often required tens or hundreds megawatts for generating high acceleration gradient, single klystron is difficult to meet the requirements or at least not the most economical solution as the pulse width of klystron is several times longer than the filling time of the accelerator structure and the duration of the electron beam, an effective way is power combining and pulse compression technology based on the moderate power klystrons [1].

In the design of RF systems for NLC at SLAC, four 50 MW X-band klystrons with pulse widths 1.6 μs were combined and pulse compressed by SLED-Ⅱ system, the output peak power was about 580 MW with pulse widths 400 ns [2]. In the design of RF systems for the C-band linear collider at KEK, two 50 MW C-band klystrons with pulse widths 2.5 μs were combined and pulse compressed by a multi-cell coupled-cavity system, the output peak power was about 350 MW with pulse widths 500 ns [3].The high power pulses generated by pulse compressors were divided and delivered into multi-sections of the accelerator structures, and the numbers of klystrons could be significantly reduced.

High power microwave systems based on power combining and pulse compression of conventional klystrons have characteristics of high peak power, good stability and long-term operation with high repetition frequency. In addition to be widely used in large scale particle accelerators, this type of microwave system could also be used as high power radar transmitters or high level electromagnetic environment generators while the output power is radiated by antennas [4][5]. Due to the high energy efficiency of this system, the demand for primary energy is relatively low. This high power microwave system is compact and easy to move, suitable for operation outdoor. In the microwave test facility designed by the US Company TITAN Beta which is operating for high power radiated susceptibility testing on military and civilian equipment, 20 MW S-band klystron and SLED type pulse compressor were used, the output peak power was about 140 MW with pulse widths 400 ns [6].

In this paper, a compact and stable high power microwave system based on power combining and pulse compression of conventional klystrons had been designed and built. The output peak power is more than 210 MW with pulse widths 400 ns at 25 Hz.

## 2. System design

The scheme of the high power microwave system is as shown in Fig.1. Two 50 MW S-band klystrons with pulse widths 4 μs are combined by a four ports power combiner, and the input power of SLED pulse compressor is about 100 MW. Assuming the peak power gain of SLED pulse compressor is 5 and the energy efficiency is 60%, the output peak power will be about 500 MW with pulse widths 480 ns.

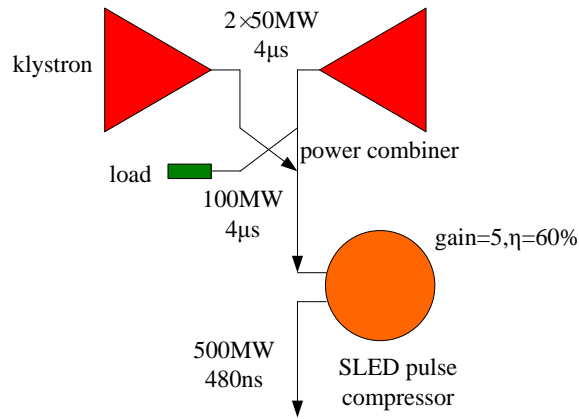

Fig.1. Schematic of the high power microwave system

### 2.1 Pulse Modulator

Pulse modulator supplies the voltage and current for the klystron. A solid state induction modulator and pulse transformer are used to drive two klystrons in parallel which requiring 325 kV 800 A pulses for 4 μs flat top at 25 Hz. The primary of pulse transformer is composed of 20 modulation units (including 2 backup units), as shown in Fig.2. The output pulse of each modulation units is about 2 kV 9.6 kA which is realized by using a single-turn transformer driven by 4 parallel IGBT cells. The solid state induction modulator can produce a 32.5 kV 4 μs pulse at 9.6 kA, with a 1 to 10 pulse transformer, the output can develop about 325 kV at 800 A.

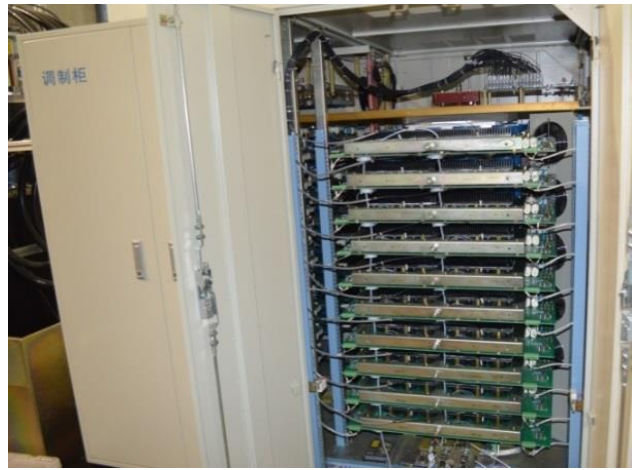

Fig.2. Photograph of the solid state induction modulator

The measured waveforms of the pulse transformer are shown as Fig.3, CH1 and CH2 are voltage waveforms, CH3 and CH4 are current waveforms. The flat top of voltage pulse is about 3.6 μs, the rising and falling edge are slower than expected.

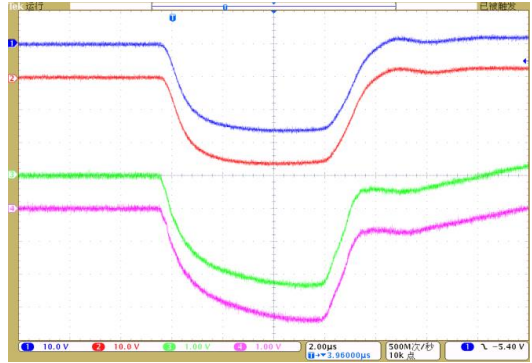

Fig.3. Classical output waveform of the pulse transformers

## 2.2 Power combiner

Power combiner is one of the key devices in this high power microwave system. For the safety of RF windows of klystrons, the reflection of each port of the power combiner should be as small as possible, while the isolation of adjacent input ports should be as high as possible, and the power capacity must be more than 100 MW.

The structure of power combiner is as shown in Fig.4, this four ports power combiner have simple structure and easy to process [7]. As only changed at H-plane, the power handling capacity of this power combiner could be raised by add the height of waveguide. When microwave pulses input simultaneously at port1 and port4 with the same frequency and amplitude, the combined pulses will output from port2 or port3 determined by the input phase difference.

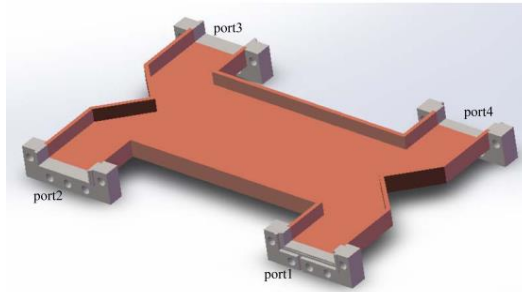

Fig.4. Sectional view of the magic H type power combiner

The cold test scattering parameters are shown as Fig.5, The return loss of each ports and the isolation of adjacent input ports are about 37 dB at 2856 MHz. This power combiner can also be used as a 3 dB coupler with the unbalance degree is less than 0.1 dB at 2856 MHz. The experiment result of power combining with klystron driving source shows that the power combining efficiency is more than 95%.

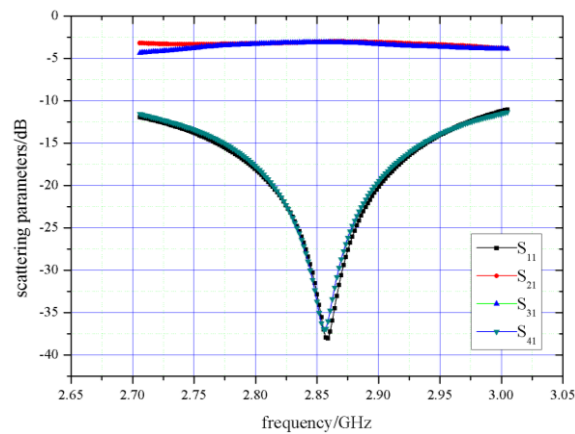

Fig.5. Measured scattering parameters of the power combiner

## 2.3 Driving source of klystrons

The driving source of klystrons in this system must meet the requirements for input microwave frequency, amplitude, and phase and time synchronization for power combining in waveguide, the requirement of phase shifting for SLED pulse compression. So a double channel solid state amplifier microwave source is designed for driving two klystrons in parallel, the block diagram of the driving source is as shown in Fig.6.

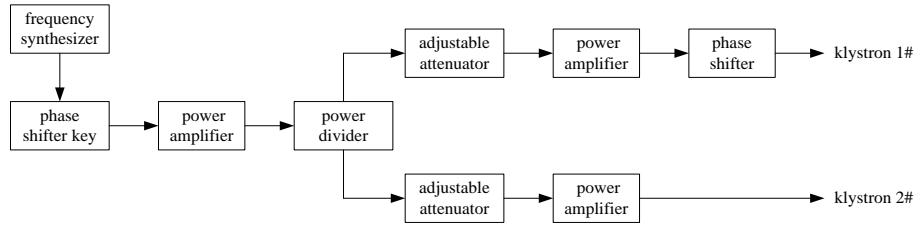

Fig.6. Schematic of the double channel solid state amplifier microwave source

A 10 mW frequency synthesizer with center frequency at 2856 MHz and a fast phase shifter key with switch speed less than 15 ns are located in the forefront of the microwave source. After the first-stage amplifier, the output peak power will be about 300 W, then the power is divided and amplified again, the output power of the driving source can be reached about 600 W. In our design, single frequency synthesizer can ensure the same frequency of two klystrons, adjustable attenuators before the second-stage amplifier are used to adjust the output power of klystrons, and a phase shifter will meet the phase difference requirement for power combining.

## 2.4 Pulse compressor

RF pulse compressor converts long-duration moderate-power input pulses into short-duration high-power output pulses. SLED type pulse compressor is used in this high power microwave system for its simple and compact structure, and the convenient way for energy extraction [8].

The SLED type pulse compressor is shown as Fig.7, it consists of a 3 dB coupler and a pair of $TE_{015}$ cylindrical storage cavities. In order to deal with the RF breakdown problem with 100 MW maximum input power, the dual side-wall coupling irises structure was used [9]. The unloaded quality factor of each cavity is about $10^5$ and the coupling coefficient is about 5. According to the theory of the SLED pulse compression, when the input pulse width is 3.6 μs and the phase changed at 3.0 μs, the normalized power is shown as Fig.8. The maximum peak power gain is about 5.5 at the time instant when the phase is reversed and then the output pulse decays exponentially.

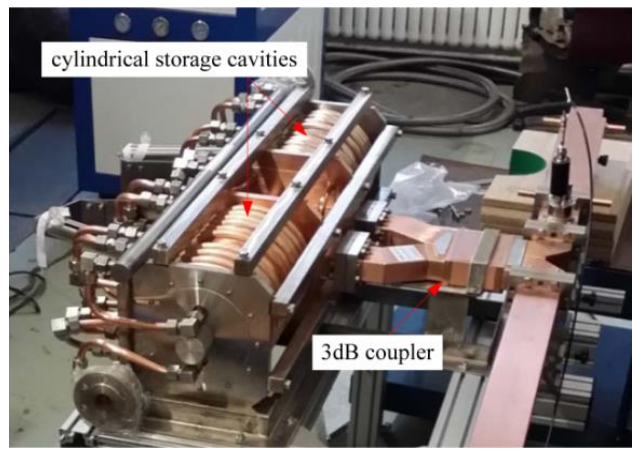

Fig.7. Photograph of the S-band SLED type pulse compressor

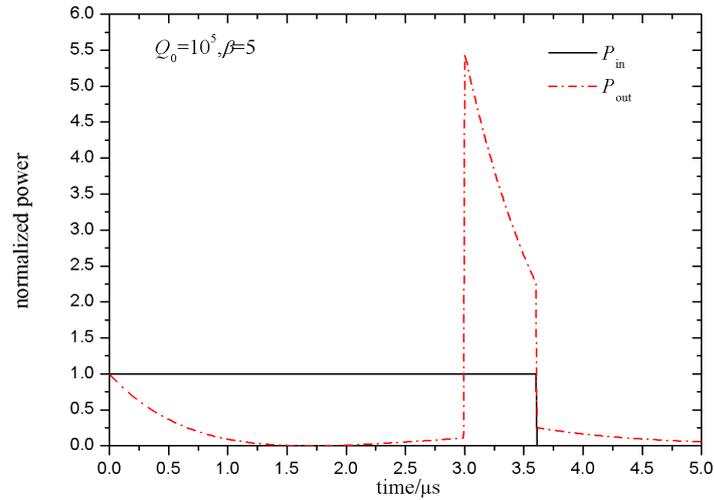

Fig.8. Normalized power of the S-band SLED pulse compressor

## 3. Experiments

The layout of klystrons and waveguide components of the high power microwave system are shown as Fig.9, directional couplers are used to monitor the power of klystrons, power combiner and SLED pulse compressor. In order to ensure the safe and stable operation of the high power microwave system, a control and protection system based on the PLC and Labview was designed for process control, condition monitor and interlock [10].

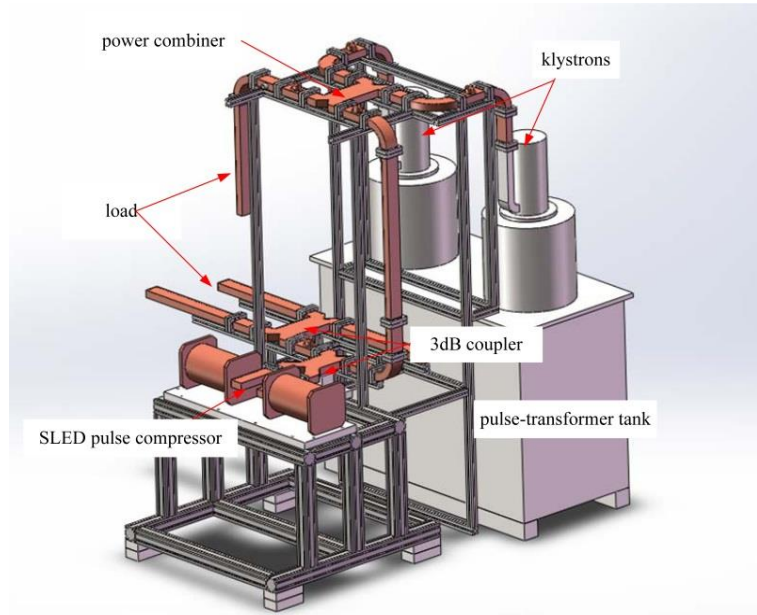

Fig.9. Layout of klystrons and waveguide components

Considering the power capacity of single load, another 3 dB power coupler is located at the output port of the SLED pulse compressor and then the output high power will be absorbed by two loads. The output peak power is controlled at about 200MW by adjusting the power of driving source of klystrons and the voltage of modulator.

The photograph of the high power microwave system based on power combining and pulse compression of conventional klystrons is shown as Fig.10, the whole system is very compact. After RF conditioning, a typical envelope waveforms of the high power microwave system are shown as Fig.11, CH1 and CH2 are the output waveforms of klystron 1# and 2#, CH3 is the output waveform of power combiner, CH4 is the output waveform of the SLED pulse compressor. When the output power of klystrons are about 22.2 MW and 23.4 MW with pulse widths 3.6 μs, the combined power is about 44.6 MW, the combining efficiency is about 97.7%. While the phase reversed at 3.0 μs, the output peak power is about 212.8 MW with pulse widths nearly 400 ns, the peak power gain is about 4.7.

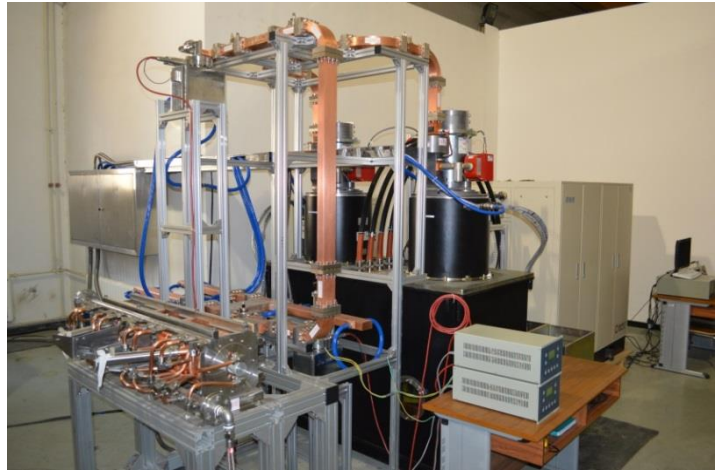

Fig.10. Photograph of the high power microwave system

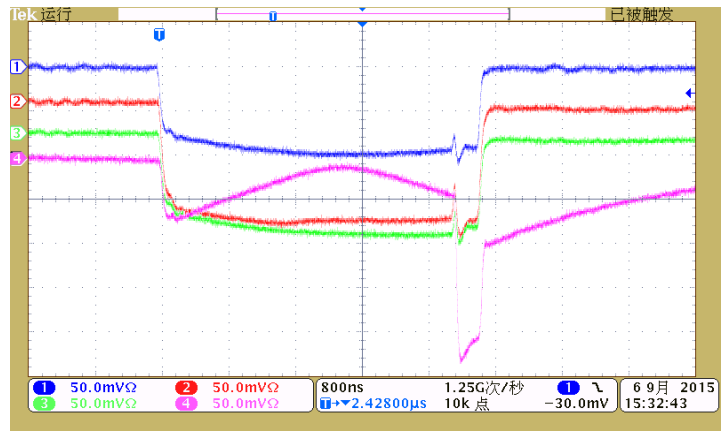

Fig.11. Typical waveforms of the high power microwave system

## 4. Conclusions

A S-band high power microwave system based on power combining and pulse compression of conventional klystrons was designed and built. The output power of two klystrons was combined with high efficiency, the compressed pulse peak power had reached more than 200 MW with pulse widths nearly 400 ns at 25 Hz. GW levels high power microwave could be generated if more klystrons were combined or several systems described above as a unit were recombined. This type of high power microwave system has the property of compact, stable operation and convenient to adjust the power and pulse width, which will has widely application prospect in RF system of large scale particle accelerators, high power radar transmitters and high level electromagnetic environment generators.

**Acknowledgements**

The author appreciates the guidance and help of researcher Feng-Li Zhao at IHEP.